\pdfoutput=1

\documentclass[11pt]{article}
\usepackage{acl}
\usepackage{times}
\usepackage{latexsym}
\usepackage[T1]{fontenc}
\usepackage[utf8]{inputenc}

\usepackage{microtype}

\usepackage{multirow}
\usepackage{graphicx}
\usepackage{comment}
\usepackage{amsmath} % for \begin{align}
\usepackage{amsfonts}
\usepackage{bbm} % for indicator variable, \mathbbm{1}
\usepackage{makecell} % \Xhline (thick line)
\usepackage{subcaption} % \subtable
\usepackage{booktabs} % for table presentation .. \toprule, \midrule, \bottomrule
\usepackage{multirow}
\usepackage{bbding} % \Checkmark \CheckmarkBold
\usepackage{pifont}% http://ctan.org/pkg/pifont
\newcommand{\cmark}{\ding{52}}%
\newcommand{\xmark}{\ding{56}}%

\title{Collective Relevance Labeling for Passage Retrieval}
\author{Jihyuk Kim \\ Yonsei University \\ \texttt{jihyukkim@yonsei.ac.kr} \\
  \And
  Minsoo Kim \\Seoul National University \\\texttt{minsoo9574@snu.ac.kr} \\
  \AND
  Seung-won Hwang\thanks{~~Corresponding author.} \\Seoul National University \\\texttt{seungwonh@snu.ac.kr}
}

\begin{document}
\maketitle
\begin{abstract}
Deep learning for Information Retrieval (IR) requires a large amount of high-quality query-document relevance labels, but such labels are inherently sparse. Label smoothing redistributes some observed probability mass over unobserved instances, often uniformly, uninformed of the true distribution. In contrast, we propose knowledge distillation for informed labeling, without incurring high computation overheads at evaluation time. Our contribution is designing a simple but efficient teacher model which utilizes collective knowledge, to outperform state-of-the-arts distilled from a more complex teacher model. Specifically, we train up to $\times8$ faster than the state-of-the-art teacher, while distilling the rankings better. 
Our code is publicly available at \url{https://github.com/jihyukkim-nlp/CollectiveKD}.
\end{abstract}

\section{Introduction}
\label{sec:introduction}

Facilitated by recent developments in pre-trained language models (PLM) such as BERT~\cite{devlin-etal-2019-bert}, neural ranking models have seen significant improvements in effectiveness~\cite{yates2021pretrained}. Neural ranking models can be categorized into two major groups: Cross-encoders and bi-encoders, which jointly or separately encode a query $q$ and a passage $p$, respectively. While the former shows stronger performance, the latter makes the passage representations indexable and thus enables efficient retrieval supported by approximated nearest neighbor (ANN) search, e.g., FAISS~\cite{johnson2019billion}. In this work, targeting web-scale retrieval, we follow the bi-encoder design and adopt the state-of-the-art bi-encoder, ColBERT~\cite{khattab2020colbert} as our target retriever.

While label supervision has a critical role in training bi-encoders, only \textbf{incomplete labels} are available in benchmark training datasets due to the prohibitive cost of exhaustive human annotation of large-scale passage corpora. As a prominent example, in MSMARCO~\cite{nguyen2016ms} training dataset, the number of labeled relevant passages per query averages only 1.1, among 8.8M passages. In contrast, in TREC-DL 2019~\cite{craswell2020overview} which provides complete annotations for a select number of queries on the same passage collection, the same value is 58.2, indicating that significant amounts of relevant passages can remain unlabeled in MSMARCO. Meanwhile, we stress that the problem of incomplete labels observed from benchmark datasets gets worse in real-life retrieval tasks, where new relevant documents are constantly added without annotation.

\begin{table}[t]
    \centering
    \resizebox{\columnwidth}{!}{
    \begin{tabular}{l|lc}
        \toprule
        Training strategy & KD capacity & Efficiency \\
        \midrule
        Standard KD & \cmark (cross-encoder) & \xmark \\
        Self-KD & \xmark (bi-encoder) & \cmark \\
        \midrule
        Ours && \\
        \quad\textbf{Collective Self-KD} & \cmark (collective bi-enc)& \CheckmarkBold \\
        \bottomrule
    \end{tabular}
    } 
\caption{Comparison of existing KD approaches with ours. Teacher for each method denoted in parentheses.}
\label{tab:informed_labeling}
\end{table}

Existing work addressing incomplete labels can be categorized into uninformed smoothing and informed labeling. \textbf{Uninformed smoothing} redistributes the observed probability mass uniformly over the unobserved, and has been shown to improve model calibration and prediction~\cite{szegedy_label_smoothing,muller_label_smoothing}. However, given diverse topics in the passage corpora, uniform smoothing does not accurately reflect the true distribution. Alternatively, \textbf{informed labeling} distills knowledge from a trained model, evaluating the relevance of a $q$-$p$ pair, to assign a higher probability to an unlabeled pair with a higher estimated relevance, also known as knowledge distillation (KD)~\cite{hinton2015distilling}.

Table~\ref{tab:informed_labeling} contrasts two existing KD strategies with different teachers -- cross- and bi-encoders -- and positions our proposed KD approach. Standard KD approaches adopt a high-capacity teacher, e.g., cross-encoder, to teach a bi-encoder student, e.g., ColBERT~\cite{hofstatter2020improving}. However, the cross-encoder teacher sacrifices efficiency, requiring BERT encodings to be recomputed for each possible $q$-$p$ pair. Alternatively, student and teacher can have equal capacity~\cite{furlanello2018born}, by adopting a bi-encoder teacher, i.e., in a self-knowledge distillation setting (self-KD). However, while increasing efficiency due to independent encoding of $q$ and $p$, the bi-encoder teacher provides little additional knowledge to the student. Our contribution is devising a more informed bi-encoder teacher achieving both high capacity and efficiency.

\begin{table}[t!]
    \centering
    \resizebox{\columnwidth}{!}{
    \begin{tabular}{l|l}
        \toprule
        $q$ & Is \textbf{caffeine} a \textbf{narcotic}? \\
        \toprule
        $p_1$ 
        & An opioid is sometimes called a \\
        (relevant) & \textbf{narcotic}. The \textit{combination} of aspirin, \\
        & \textit{butalbital}, \textbf{caffeine}, and codeine is \\
        & used to \textit{treat tension headaches}. ...\\
        \midrule
        $p_2$
        & The \textit{combination} of acetaminophen, \\
        (relevant) & \textit{butalbital}, and \textbf{caffeine} is used to \\
        & \textit{treat tension headaches} ...\\
        \midrule
        $p_3$ 
        & ... \textbf{Caffeine} is a considered a safe \\
        (non-relevant) & ingredient. It is stimulant that excites \\
        & the nerve cells of the brain ... \\
        \bottomrule
        $\tilde{q}$ (collective & \textit{treat tension headaches} \\
        ~~~~~knowledge) & | \textit{combination} | \textit{butalbital} \\
        \bottomrule
    \end{tabular}
    } 
\caption{A query $q$ with top-3 passages $p_{[1/2/3]}$ ranked by ColBERT. 
Key query terms are denoted in bold, 
and collective centroids
$\tilde{q}$ of $p_{[1/2/3]}$ are italicized.}
\label{tab:example}
\end{table}

To augment capacity without sacrificing efficiency, our distinction is utilizing a set of relevant passages from the bi-encoder teacher as \textbf{collective knowledge}. To illustrate, in Table~\ref{tab:example}, we show a query $q$, along with a set of three passages $\mathcal{F}_q=\{p_1, p_2, p_3 \}$, labeled as relevant by ColBERT. Considered individually, each passage may be mistakenly judged as relevant to the query, e.g., $p_3$. Instead, we consider $\mathcal{F}_q$ collectively, identifying their \textit{representative semantics}, or collective centroid $\tilde{q}$. Next, we treat $q\cup \tilde{q}$ as if it were a query, and \textit{redistribute} relevance labels based on the similarity to $q\cup \tilde{q}$, such that relevant passages (e.g., $\{p_1,p_2\}$) and non-relevant passages (e.g., $\{p_3\}$) can be discriminated. We henceforth denote our ColBERT teacher leveraging collective knowledge as \textbf{collective bi-encoder}.

We validate the effectiveness of our proposed collective bi-encoder for KD, comparing retrieval efficacy with existing models on the TREC and MSMARCO datasets. While achieving the state-of-the-art performance, we train up to $\times8$ times faster than existing KD that uses cross-encoder teachers.

\section{Approach}\label{sec:approach}
Taking the state-of-the-art bi-encoder ColBERT as our target retriever (\S\ref{sec:colbert}), our goal is to tackle the problem of incomplete labels during training (\S\ref{sec:training}). To do so, we devise a stronger ColBERT teacher that leverages collective knowledge to refine relevance labels that are transferred to the ColBERT student via KD (\S\ref{sec:teacher}).

\subsection{Baseline: ColBERT}\label{sec:colbert}
While following a bi-encoder design to enable $p$ to be indexed offline,
ColBERT additionally models term-level interactions between separately encoded $q$, $p$ term representations (called \textbf{late-interaction}) to leverage exact-match signals as in cross-encoder.

Specifically, terms in $q$, $p$ are first encoded using BERT, as contextualized representations $\mathbf{q}, \mathbf{p}\in\mathbb{R}^{768}$,
then $q$-$p$ relevance $\phi_q(p)$ is computed by aggregating similarity between the terms in $q$, $p$:
\begin{align}
    \phi_q(p)&=
    \sum_{i\in [1,|q|]} 
    \max_{j\in [1,|p|]}
    {
        (\mathbf{W}\mathbf{q}_{i})^\top
        (\mathbf{W}\mathbf{p}_{j})
    }
    \label{eq:maxsim},
\end{align}
where $|q|$, $|p|$ denote the number of tokens in $q$, $p$ respectively, and $\mathbf{W}\in\mathbb{R}^{128\times 768}$ compresses features for efficient inner-product computation.

Our goal is to improve training by addressing the challenge of incomplete labels, which we discuss in the following subsection.

\subsection{Challenge: Incomplete Labels}\label{sec:training}
We formally revisit the challenge of incomplete labels in IR. Given complete labels on true relevance, $\phi_q^*$, the correlation between $\phi_q$ and $\phi_q^*$ can be increased via an objective $\mathcal{L}$ using KL-divergence. 
\begin{align}
    \mathcal{R}_q(p)&=\text{softmax}(\phi_q(p))\\
    \mathcal{L}&=-\sum_{p'\in\mathcal{P}}\mathcal{R}^*_q(p')\log\dfrac{\mathcal{R}_q(p')}{\mathcal{R}^*_q(p')},
\end{align}
where $\mathcal{R}_q^*$ denotes a probability distribution obtained by normalizing $\phi_q^*$. Let us now consider the scenario of incomplete label supervision: The standard training approach would approximate $\mathcal{R}_q^*$ as $\mathcal{R}^*_q(p^+)$ = $1$ for a labeled positive $p^+$, and $\mathcal{R}^*_q(p^-)$ = $0$ for the others by assuming them to be negatives $\{p^-\}$:
\begin{align}
    \mathcal{L}&=-\log \mathcal{R}_q(p^+)\\
    &=-\log \dfrac{e^{\phi_q(p^+)}}{e^{\phi_q(p^+)}+\sum_{p^-}e^{\phi_q(p^-)}}\label{eq:hard_loss},
\end{align}
where, in the official train samples of MSMARCO, $p^-$ is randomly sampled from the top-1000 ranked passages by BM25 given $q$ as a query. Our claim is that, in the case of \textit{incomplete} labels, where the number of unlabeled $p^+$ becomes larger, this assumption causes $\mathcal{R}_q^*$ to be poorly approximated.

\subsection{Proposed: Collective Self-KD}\label{sec:teacher}
To better estimate $\mathcal{R}_q^*(p)$, we utilize collective knowledge, introduced in Section~\S\ref{sec:introduction}. Collective knowledge makes it possible to distill residual relevance knowledge between teacher and student, leading to a high degree of efficiency.

As illustrated in Table~\ref{tab:example}, our collective bi-encoder teacher first obtains the collective centroid $\tilde{q}$ from top-ranked passages to expand $q$ to $q\cup \tilde{q}$. Given $q\cup \tilde{q}$, the collective knowledge of the teacher is reflected in the updated $\mathcal{R}_q^*(p)$, which is now approximated as $\mathcal{R}_{q\cup \tilde{q}}(p)$, the similarity between $p$ and $q\cup \tilde{q}$. In practice, we implement our teacher using the recently proposed ColBERT-PRF~\cite{10.1145/3471158.3472250}, which uses \textbf{PRF for \textit{query expansion}} (at test time) by using top-ranked passages from ColBERT as pseudo-relevance feedback (PRF).

Our distinction is leveraging \textbf{PRF for \textit{labeling}} (at train time) to augment the capacity of bi-encoder teacher. This is an important distinction, as it provides substantial benefits as follows: (1) Train-time PRF is more reliable than test-time PRF, since at test time, the initial ranking is less accurate for novel queries that have not been observed during training. (2) Train-time PRF additionally provides hard negatives~\cite{xiong2021approximate} (e.g., $p_3$ in Table~\ref{tab:example}), as by-products of the initial ranking from ColBERT, producing better $\{p^-\}$ for training. With the goal of distilling collective knowledge while preserving test time efficiency, we now describe in detail the operation of our collective bi-encoder teacher, and $\mathcal{R}_q^*(p)$ estimation process. 

\subsubsection{Collective Knowledge Extraction}
As a preliminary, we pre-train a ColBERT model using Eq (\ref{eq:hard_loss}) and use it to rank passages in the collection for each $q$, producing a ranking $\Pi_q=\{p^l\}_{l=1}^{|\mathcal{P}|}$ where $l$ and $|\mathcal{P}|$ denotes the rank of each passage and the size of the passage collection, respectively. From these, we obtain the top-$f_p$ ranked passages, as feedback passages $\Pi_q^{l\le f_p}=\{p^l\}_{l=1}^{f_p}$.

To extract collective knowledge present in $\Pi_q^{l\le f_p}$, we apply k-means clustering on the token embeddings in $\Pi_q^{l\le f_p}$, i.e., $\{\mathbf{W}\mathbf{p}^l_i\}_{\forall l\in\{1,f_p\}, \forall i\in\{1, |p^l|\}}$, producing $f_c$ centroid embeddings $\mathcal{C}_{q}=\{\mathbf{c}_{q,m}\}_{m=1}^{f_c}$. Then, in order to sift out trivial knowledge, discriminative embeddings among $\mathcal{C}_{q}$ are further filtered. More precisely, we sort the centroids in $\mathcal{C}_{q}$, by the IDF score of the token that is the nearest neighbor, among the entire passage collection, to each $\mathbf{c}_{q,m}$.  According to these IDF scores, we select the top-$f_e$ embeddings among $\mathcal{C}_{q}$ as the final collective centroid embeddings, i.e., vector representations for $\tilde{q}$, denoted by $\mathcal{E}_q=\{\mathbf{e}_{q,n}\}_{n=1}^{f_e}\subset \mathcal{C}_{q} $.

Finally, to obtain the improved approximation of $\mathcal{R}_q^*$ via $\mathcal{R}_{q\cup \tilde{q}}(p)$, the similarity between $\mathcal{E}_q$ and $p$ is added to $\phi_q(p)$ in Eq (\ref{eq:maxsim}):
\begin{align}
    \begin{split}
    \mathcal{R}^*_q(p)&\approx\text{softmax}( \phi_q(p) \\
    &+\beta \sum_{n\in[1,f_e]}\sigma_{q,n}\max_{j\in [1,|p|]}{\mathbf{e}_{q,n}^\top (\mathbf{W}\mathbf{p}_{j}})),
    \end{split}
    \label{eq:teacher_score}
\end{align}
where $\sigma_{q,n}$ is the IDF of the nearest neighbor token of $\mathbf{e}_{q,n}$, weighing the contribution of each $\mathbf{e}_{q,n}$ in terms of discriminability, and $\beta$ is a hyper-parameter controlling overall contribution. Note that, we can reuse the collective centroids for estimating the $\mathcal{R}_q^*(p)$ of any $p$, avoiding redundant BERT encoding. 

\subsubsection{Collective Knowledge Distillation}
We now propose efficient distillation strategies for teaching $\mathcal{R}_q^*$ to the student, where we leverage the fact that the student and the teacher share the identical architecture design. For efficient KD, we let our student inherit most of the knowledge through the parameters of the teacher, by initializing the student's parameters with those of the pre-trained ColBERT, denoted by $\theta$. Since $\theta$ already captures $\mathcal{R}_q$, the knowledge distillation of $\mathcal{R}_{q\cup \tilde{q}}$ can be simplified as distilling residual relevance, i.e.,  $\mathcal{R}_{\tilde{q}}$, to the student.

Meanwhile, when using $\theta$ for initialization, $p^-$ sampled by BM25 now becomes a trivial negative that can be easily discriminated by $\theta$. To learn additional knowledge, we introduce hard negatives, by utilizing the initial ranking $\Pi_q$ from the pre-trained ColBERT. Specifically, we replace the pool of $p^-$ with top-100 ranked passages $\Pi_q^{l\le 100}$, so as to improve top-ranking performance of the student.

\begin{table*}[ht]
    \centering
    \resizebox{\textwidth}{!}{
    \begin{tabular}{l|cc|cc|cc|c}
        \toprule
        \multirow{2}{*}{Retriever} 
        & \multicolumn{2}{c|}{MSMARCO Dev}            & \multicolumn{2}{c|}{TREC Passages 2019} & \multicolumn{2}{c|}{TREC Passages 2020} &  MRT \\
        \cline{2-7}
        & MRR@10 & Recall@1k & NDCG@10 & Recall@1k & NDCG@10 & Recall@1k &  (ms)   \\
        \midrule
ColBERT-PRF Ranker ($\beta=0.5$) & 0.334 & 0.974 & 0.776 & 0.899 & 0.732 & 0.910 & \multirow{2}{*}{3035} \\
ColBERT-PRF Ranker ($\beta=1.0$) & 0.317 & 0.965 & 0.766 & 0.885 & 0.704 & 0.900 &  \\ 
\midrule
ColBERT (a) & 0.367 & 0.967 & 0.700 & 0.837 & 0.675 & 0.854 & \multirow{7}{*}{1299} \\
ColBERT-HN (b) & 0.374 & 0.969 & 0.741 & 0.838 & 0.700 & 0.867 & \\
Self-KD (c) & 0.367 & 0.968 & 0.705 & 0.833 & 0.704 & 0.839 & \\
Standard KD; CE (d) & 0.373 & 0.967 & 0.715 & 0.791 & 0.696 & 0.829 & \\
Standard KD; CE-3 (e) & 0.378 & 0.963 & 0.717 & 0.793 & 0.684 & 0.824 & \\
\cline{1-7}
(ours) Collective Self-KD ($\beta=0.5$) 
& \textbf{0.386}$^{a,b,c,d}$ & \underline{0.972}$^{a,b,c,d,e}$ & \underline{0.744} & \underline{0.843} & \textbf{0.724} & \underline{0.874}$^{e}$ &  \\
(ours) Collective Self-KD ($\beta=1.0$) 
& \underline{0.384}$^{a,c,d}$ & \textbf{0.974}$^{a,b,c,d,e}$ & \textbf{0.751} & \textbf{0.868} & \underline{0.721} & \textbf{0.883}$^{e}$ &  \\
        \bottomrule
    \end{tabular}
    }
    \caption{
    ``CE'' and ``CE-3'' denote a single and an ensemble of three cross-encoder teachers, respectively.
    MRT denotes Mean Response Time (ms) per query. 
    Those outperforming all baselines except ColBERT-PRF are  \underline{underlined}, and the best shown in \textbf{bold}.
    Superscripts denote statistical significance with paired t-test over the indicated baselines ($p<0.05$). 
    }
\label{tab:student_ranking}
\end{table*}

\section{Experiments}
\subsection{Experimental Settings}\label{sec:exp_setup}
We conduct our experiments by trainining retrievers using the MSMARCO training dataset, and evaluating on the MSMARCO Dev and TREC-DL 2019/2020 datasets, which provide binary and graded relevance labels, respectively.
For TREC-DL datasets, we adopt grade 2 as the binary cut-off.
For evaluation metrics, we report MRR@10 for MSMARCO and NDCG@10 for TREC-DL, and Recall@1000 for both. We also report the per query mean response time (MRT) of retrievers. The results are from a single run.

\subsection{Dataset Details}
\label{sec:data_details}
MSMARCO~\cite{nguyen2016ms}\footnote{MSMARCO is intended for non-commercial research purposes. Refer to https://microsoft.github.io/msmarco for further details.} is a passage ranking dataset initially introduced for reading comprehension and subsequently adopted for retrieval. The collection consists of 8.8M passages, obtained from the top-10 search results retrieved by the Bing search engine, in response to 1M real-world queries. The training and evaluation sets contain approximately 500,000 queries and 6,980 queries, respectively, with roughly one relevant positive passage label per query, with binary relevance labels. 

TREC-DL 2019/2020~\cite{craswell2020overview,craswell2021overview} datasets are provided by the passage ranking task of the 2019 and 2020 TREC Deep Learning tracks, providing densely-judged annotations of 43 and 54 queries, respectively. They share the same passage pool as MSMARCO, and there are 215/211 human relevance annotations per query. The relevance judgments are graded on a four-point scale: Irrelevant, Related, Highly Relevant, and Perfectly Relevant.

\subsection{Results}\label{exp:contribution_kd}
In this section, we validate the effectiveness of our collective self-KD with ColBERT as our target retriever. The results of ranking experiments are reported in Table~\ref{tab:student_ranking}.

\paragraph{Baselines}
We compare several different training strategies: training using incomplete labels (i.e., by Eq (\ref{eq:hard_loss})) or using better approximation of $\mathcal{R}^*$ from a teacher. For the former, we report results from different $\{p^-\}$: ColBERT trained using top-1000 ranked passages by BM25 (\textbf{ColBERT}) and ColBERT further fine-tuned using $\Pi_{q}^{l\le 100}$ (\textbf{ColBERT-HN}). For the latter, we compare different teachers: (1) an identically parameterized teacher (\textbf{Self-KD}), (2) a cross-encoder adopting BERT-Base encoder (\textbf{Standard KD; CE}), (3) an ensemble of three cross-encoders with different PLMs, i.e., BERT-Base, BERT-Large, and ALBERT-Large~\cite{albert}, (\textbf{Standard KD; CE-3}), and finally (4) our collective bi-encoder teacher (\textbf{Collective Self-KD}). For relevance labels from cross-encoder teachers, we used open-sourced data by \cite{hofstatter2020improving}\footnote{\url{https://zenodo.org/record/4068216}}. Analysis on estimated $\mathcal{R}^*$ from different teachers can be found in Appendix~\ref{sec:appendix_approx}. For values of ($f_p$, $f_c$, $f_e$, $\beta$), we set ($f_p$=3, $f_c$=24, $f_e$=10) and ran experiments with $\beta=0.5$ and $\beta=1.0$, by referring to the reported results in \citet{10.1145/3471158.3472250}. For further analysis on ($f_p$, $f_c$, $f_e$, $\beta$), see Appendix~\ref{sec:prf_configuration}. In addition, we also compare ColBERT-PRF Ranker with $\beta=0.5$ and $\beta=1.0$.

\paragraph{Ranking Performance}
By leveraging PRF to augment query contexts, ColBERT-PRF shows strong performance, outperforming the others on TREC-DL datasets. Meanwhile, on MSMARCO Dev queries, ColBERT-PRF shows comparable or higher recall but lower MRR@10, compared to ColBERT. This is a well-known limitation of PRF for sparse query datasets, such as MSMARCO evaluation dataset with a few relevant documents (1.1 per query)~\cite{amati2004query}.

Among ColBERT retrievers, as expected, basic self-KD fails to improve performance. For ColBERT students distilled from cross-encoder teachers, Recall@1k shows marginal difference on MSMARCO Dev, and decreases on the TREC-DL datasets. Our teacher is the most effective for KD, where our student outperforms the others on all metrics. As a result, our student shows closest performance to ColBERT-PRF. This indicates collective knowledge from PRF helps to better labeling relevance and our collective self-KD effectively transfers collective knowledge to our student.

Importantly, our method produces a much more efficient student than ColBERT-PRF, by distilling collective knowledge into the parameters of the student, resulting in a 3-fold reduction of MRT. More specifically, recall that ColBERT-PRF performs PRF at \textit{evaluation time}, and thus performs two rounds of retrieval (one using $q$ and the other using $q\cup \tilde{q}$), increasing latency approximately two fold. Furthermore, obtaining $\tilde{q}$ from PRF requires additional online computation at evaluation time. In contrast, we transfer such computations to training time for relevance labeling, eliminating all overheads at evaluation time.

\paragraph{Training Efficiency}\label{sec:efficient_kd}
We compare training efficiency between our collective self-KD and standard KD using cross-encoder teachers. On the same device\footnote{We used a single GeForce RTX 3090 GPU with 24 GiBs of memory, on a server which has 187 GiBs of RAM, and two Intel Xeon Gold 6254 CPUs, each with 18 physical cores.}, we measured elapsed times for the annotation phase of relevance labels via teacher, and student training phase using the labels.

In annotation phase, for a single cross-encoder teacher using a BERT-Base encoder, obtaining $\mathcal{R}^*$ takes roughly 40 hours. For a teacher using a BERT-Large encoder, this time increases to 90 hours, and for ALBERT-Large encoder, to 110 hours. When ensembling cross-encoder teachers, the cost is compounded. In contrast, our teacher takes only 15 hours for obtaining collective knowledge, i.e., $\mathcal{E}_q$ along with $\sigma_n$ in Eq~\ref{eq:teacher_score}, indicating a speedup between $\times2.67\sim\times16$. Such efficiency gain comes from the difference in the encoding phase. More precisely, given $|Q|$ queries and $|D|$ documents, cross-encoder spends quadratic complexity for encoding, e.g., $|Q|\times |D|$ BERT encoding, while our teacher adopting bi-encoder design only spends linear complexity, e.g., $|Q|+|D|$ BERT encoding. As a result, cross-encoder teachers do not scale to real-world retrieval tasks, for labeling large numbers of queries/documents.

For training student, we enable efficient training via informed initialization. As a result, time consumed for training student only takes 20 hours, whereas the same value was around 50 hours in standard KD, indicating an overall speedup of $\times2.5$. In total, our collective self-KD is at least $\times 2.5$ faster and up to $\times 8$ faster than standard KD approaches.

\section{Conclusion and Future Work}
We study collective relevance labeling to overcome incomplete labels in passage retrieval. Our approach bypasses the computational overhead associated with PRF, leading to a state-of-the-art student retriever without sacrificing efficiency. We validate the effectiveness of our method over existing approaches on the MSMARCO and TREC-DL Passage Ranking datasets. 

As future work, we consider zero-shot transfer. While our collective self-KD effectively distills knowledge, one requirement we have is to bootstrap the collective knowledge with a good initial ranking model. As it usually requires some form of label supervision to train such a model, the case where no initial supervision is available, may be considered a limitation of our method. We believe bootstrapping an effective self-KD without any supervision is a promising direction.

\section*{Acknowledgements}
This work was supported by Microsoft Research Asia and IITP [(2022-00155958, High Potential Individuals Global Training Program) and (NO.2021-0-01343, Artificial Intelligence Graduate School Program (Seoul National University)].

\bibliography{anthology,custom}
\bibliographystyle{acl_natbib}

\appendix
\section{Appendix}\label{sec:appendix}
\begin{table*}[ht]
    \centering
    \resizebox{0.9\textwidth}{!}{
    \begin{tabular}{l|ccc|cc|cc|cc}
        \toprule
        \multirow{2}{*}{Performance Metric} & \multicolumn{3}{c|}{$f_p$} & \multicolumn{2}{c|}{$f_c$} & \multicolumn{2}{c|}{$f_e$} & \multicolumn{2}{c}{$\beta$} \\
        \cline{2-10}
        & 1 & 3 & 5 & 12 & 24 & 5 & 10 & 0.5 & 1.0 \\
        \midrule
        NDCG@10 
        & 0.651 & 0.703 & 0.702     % fp=1,3,5 
        & 0.717 & 0.703             % fc=12,24
        & 0.700 & 0.703             % fe=5,10
        & 0.710 & 0.703 \\          % beta=0.5, 1.0
        Recall@1k 
        & 0.843 & 0.842 & 0.849     % fp=1,3,5 
        & 0.846 & 0.842             % fc=12,24 
        & 0.844 & 0.842             % fe=5,10 
        & 0.848 & 0.842 \\          % beta=0.5, 1.0
        \bottomrule
    \end{tabular}
    } 
    \caption{Ranking performance of our collective bi-encoder teacher under different configurations.}
\label{tab:prf_configuration}
\end{table*}

\subsection{Analysis on $\mathcal{R}^*_q$ approximated by different teachers}\label{sec:appendix_approx}
\begin{figure*}[t]
    \centering
    \includegraphics[width=\textwidth]{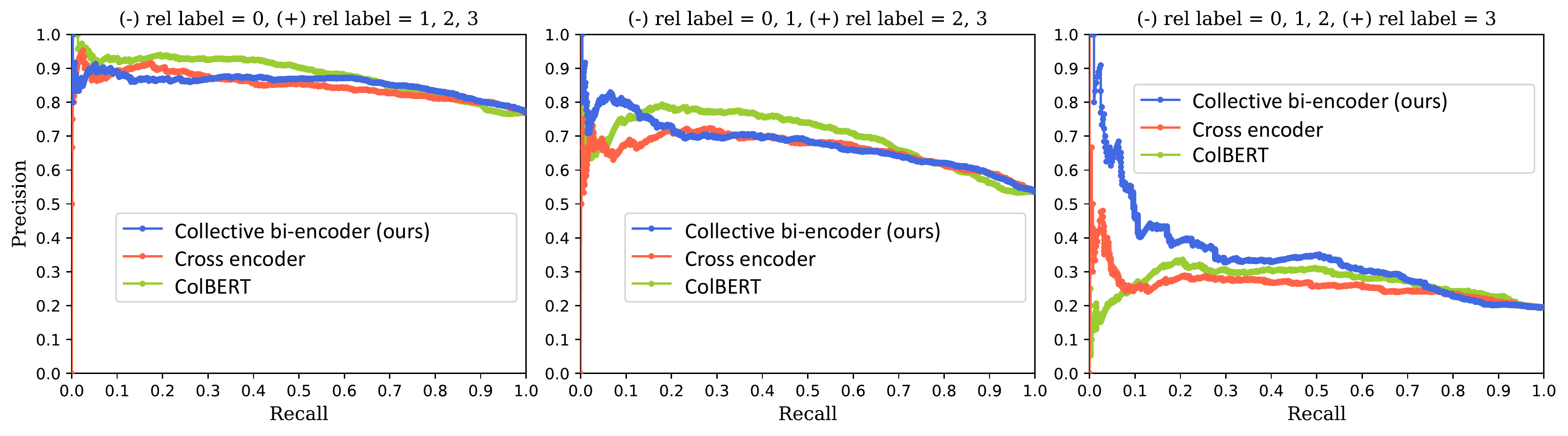}
    \caption{Precision-recall curves on TREC 2019 with different thresholds on predicted/labeled relevance scores. From the left to the right, we gradually increase the threshold of the labeled relevance that decides which passages are positive passages.}
    \label{fig:pr_curve_all}
\end{figure*}

Given the top-100 passages ranked by the pretrained ColBERT, i.e., $\Pi_q^{l\le100}$, the goal of our teacher is to obtain better relevance boundary through collective knowledge, by which unlabeled positives can be discriminated from true negatives. To analyze the reliability of relevance boundary from the teacher, we evaluate how well the teacher classifies positives among the top-100 passages, using TREC 2019 that provides complete relevance labels.

Figure~\ref{fig:pr_curve_all} shows trade-offs between precision and recall when applying different thresholds on $\phi_q$. We set different cut-offs on graded relevance for deciding positive passages. For example, in the left most figure with boundary 1, we treat passages with labeled relevance 1,2,3 as positives and 0 as negatives. As baselines, we compare our collective bi-encoder teacher with the pretrained ColBERT, and a cross-encoder teacher, monoBERT\footnote{https://github.com/castorini/pygaggle/}~\cite{nogueira2019multistage}. Better precision-recall curves should bow towards the top right corner.

For the cut-offs of 1 and 2, we observed little differences between retrievers. For example, when retrievers are tasked to classify non-relevant passages with relevance label 0 and the others (the left most figure), all retrievers show 100\% recall with near 80\% precision. In contrast, for the cut-off of 3 (the rightmost figure), our teacher shows much better trade-offs compared to the others. For example, collective bi-encoder teacher shows near 50\% precision to achieve 10\% recall, while the others show 20\% precision to achieve the same recall. The noticeable drop of precision for both ColBERT and cross-encoder in low (<0.5) recall regimes indicate that these models have difficulty distinguishing moderate relevance (1, 2) from perfect relevance (3), that is, they are not well calibrated to distinguish between those two groups. On the other hand, our teacher's refinement of $q$ through collective knowledge is effective in calibrating relevance, to finely reflect the distinction between perfectly relevant passages and the others.

\subsection{Analysis on $(f_p, f_c, f_e, \beta)$}\label{sec:prf_configuration}
Exploring ranking performance of students under all different configurations of $(f_p, f_c, f_e, \beta)$ is expensive. Instead, the optimal configuration can be decided by ranking performance of the teacher. Here, we explore  $f_p\in\{1,3,5\}$, $f_c\in\{12, 24\}$, $f_e\in\{5, 10\}$, and $\beta\in \{0.5, 1.0\}$. Meanwhile, according to reported results by \cite{10.1145/3471158.3472250}, $(f_p=3, f_c=24, f_e=10, \beta\in\{0.5, 1.0\})$ shows strong performance. Thus, instead of testing all different configurations, we set $(f_p=3, f_c=24, f_e=10, \beta=1.0)$ as default values, and change one variable at a time. When evaluating teachers using TREC-DL 2019 dataset, we found similar results to \cite{10.1145/3471158.3472250} (Table~\ref{tab:prf_configuration}).
\end{document}